\documentclass[12pt]{iopart}
\usepackage{iopams}
\usepackage[pdftex]{graphicx}
\usepackage{citesort}

\begin{document}

\title[Towards magnetic control of magnetite]{Towards magnetic control of
magnetite}
%{a detailed study on the influence of substrate, orientation, and thickness of high quality stoichiometric films.}

\author{F.~J.~Pedrosa,$^1$ J.L.F.~Cu\~{n}ado,$^{1,4}$ P. Perna,$^1$ M.~Sanz,$^2$ M.~Oujja,$^2$ E.~Rebollar,$^2$ J.F.~Marco,$^2$ J.~de la Figuera,$^2$ M.~Monti,$^2$ M.~Castillejo,$^2$ M.~Garc\'ia-Hern\'andez,$^3$ F.~Mompe\'an,$^3$ J.~Camarero,$^{1,4}$  and A.~Bollero$^1$ }
\address{$^1$ Instituto Madrile\~{n}o de Estudios Avanzados en Nanociencia, IMDEA Nanociencia, Campus Universidad Aut\'onoma de Madrid, 28049, Madrid, Spain}
\address{$^2$ Instituto de Qu\'imica F\'isica "Rocasolano", IQFR-CSIC, 28006, Madrid, Spain}
\address{$^3$ Instituto de Ciencia de Materiales, ICMM-CSIC, Campus Universidad Aut\'onoma de Madrid, 28049 Madrid, Spain}
\address{$^4$ Departamento de F{\'i}sica de la Materia Condensada and Instituto "Nicol{\'a}s Cabrera", Universidad Aut{\'o}noma de Madrid, Campus Universidad Aut\'onoma de Madrid, 28049 Madrid, Spain}

\ead{javier.pedrosa@imdea.org}
\ead{alberto.bollero@imdea.org}
\vspace{10pt}

\begin{indented}

\item[]\today
\end{indented}

\begin{abstract}

High quality stoichiometric magnetite (Fe$_3$O$_4$) films grown by
infrared pulsed laser deposition (IR-PLD) on different surfaces
have been investigated in order to study the influence of the
substrate, orientation, and thickness on their magnetic behavior.
Different single crystal (001)-oriented substrates, i.e.,
SrTiO$_3$(001), MgAl$_2$O$_4$(001)and MgO(001), have been used for
the preparation of epitaxial Fe$_3$O$_4$(001) films. By
comparison, polycrystalline magnetite films were obtained on both
single crystal Al$_2$O$_3$(0001) and amorphous Si/SiO$_2$
substrates. The thickness has been varied between 50 - 400 nm. All
films consist of nanocrystalline stoichiometric magnetite with
very small strain ($<1\%$) and present the Verwey transition
($T_{\rm V}$) between 110-120 K, i.e., close to bulk magnetite
(122 K). In general, $T_{\rm V}$ depends on both microstructure
and thickness, increasing mainly as the thickness increases. Room
temperature angular-dependent measurements reveal an in-plane
fourfold symmetry magnetic behavior for all films grown on
(001)-oriented surfaces, and with the easy axes lying along the
Fe$_3$O$_4$[010] and [100] directions. Remarkably, the fourfold
magnetic symmetry shows up to 400 nm thick films. In turn, the
films grown on single crystal Al$_2$O$_3$(0001) and on amorphous
Si/SiO$_2$ surfaces display an isotropic magnetic behavior. In
general, the coercive field ($H_{\rm C}$) depends on
microstructure and film thickness. The largest (lowest) $H_{\rm
C}$ value has been found for the thinner film grown on a single
crystal SrTiO$_3$(001) (amorphous Si/SiO$_2$) surface, which
present the largest (lowest) strain (crystallinity). Moreover, the
coercivity follows an inverse law with film thickness. Our results
demonstrate that we can artificially control the magnetic behavior
of stoichiometric IR-PLD grown Fe$_3$O$_4$ films by exploiting
substrate-induced anisotropy and thickness-controlled coercivity,
that might be relevant to incorporate magnetite in future
spintronic devices.
\end{abstract}

%\noindent{\it Keywords}: magnetite, thin films, magnetic anisotropy, reversal
%\submitto{\NJP}
%\maketitle

\section{Introduction}
\begin{figure}[tp]
    \begin{center}
        \includegraphics[width=0.95\textwidth]{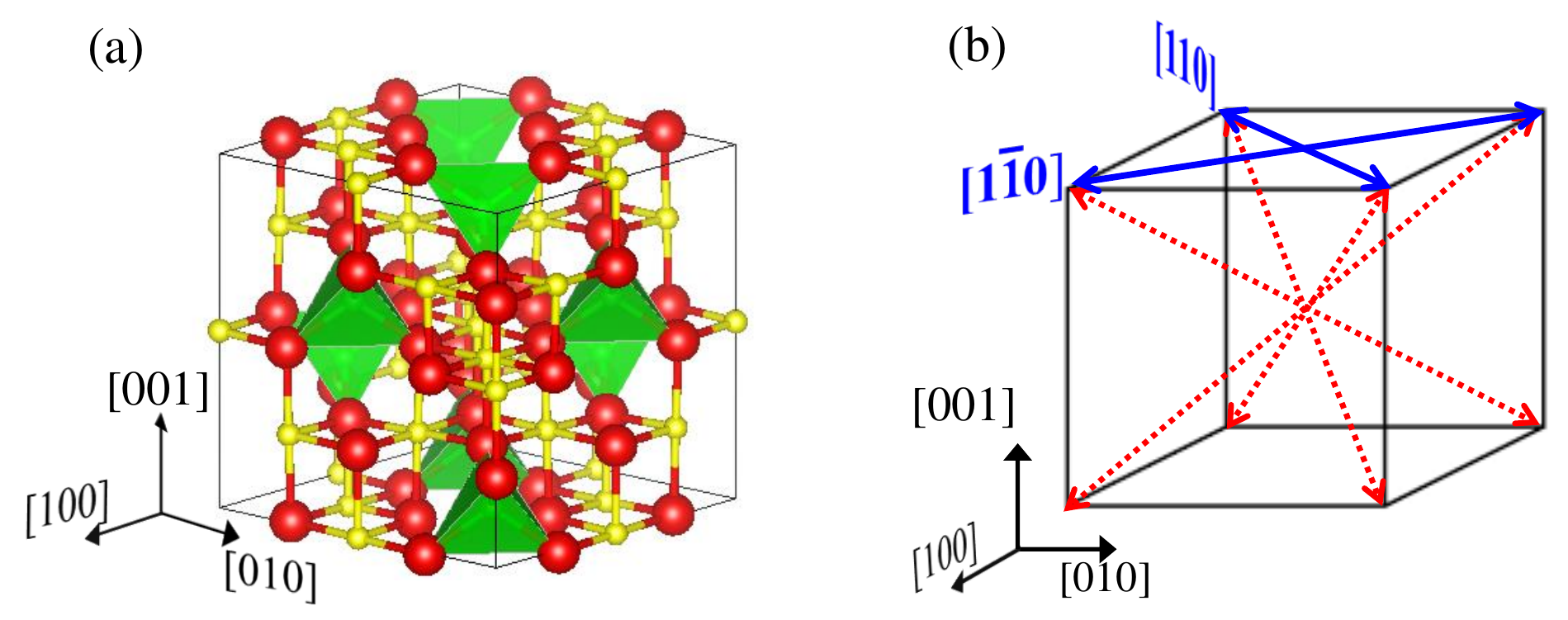}
    \end{center}
    \caption{Schematic illustration showing the crystal unit cell
    (a) and the magnetic symmetry of Fe$_3$O$_4$(001) single
    crystal at room temperature. The crystal scheme has been prepared by VESTA~\cite{VESTA}, with tetrahedral and octahedral sites
    of the inverse spinel structure indicated in (a). The
    eight anisotropy directions of bulk magnetite, i.e.,
    $\langle111\rangle$ crystal directions, are indicated with
    dashed red vectors in (b), which results with a fourfold
    in-plane magnetic symmetry in the surface plane with easy axis
    directions along the $\langle110\rangle$ surface crystal
    directions, indicated with blue vectors.}
    \label{fig:bulkMagnetite}
\end{figure}

Magnetite (Fe$_3$O$_4$) is a ubiquitous iron oxide mineral that is
the most magnetic of all the naturally-occurring minerals on
Earth~\cite{mills2004lodestone}. In the last decades, artificially
grown magnetite films have shown great potential for spintronic
applications~\cite{van2000construction,ziese2002extrinsic,wada2010efficient},
due to its robust ferrimagnetism down to nanometer thickness, high
Curie temperature (858 K), good electrical conductivity and
presumed half-metal
character~\cite{mills2004lodestone,fonin2007magnetite}, which
require well-defined and controlled magnetic behavior. Fe$_3$O$_4$
presents an inverse spinel crystalline structure at room temperature
(Fig.~\ref{fig:bulkMagnetite}.a), with a third of Fe atoms
(Fe$^{\rm 3+}$) occupying the tetrahedral A-sites and the rest (Fe$^{\rm 2+}$
and Fe$^{\rm 3+}$) on the octahedral B-sites. It presents a low-temperature
metal-insulator transition known as the Verwey
transition ($T_{\rm V}$)~\cite{walz2002verwey,garcia2004verwey},
whose temperature and character depends on crystal quality~\cite{aragon1985influence,liu2014verwey},
where the crystalline structure changes from cubic to monoclinic.
Different routes have been used to prepare magnetite thin films,
such as molecular beam epitaxy
(MBE)~\cite{Margulies1996MBE,van1998ferromagnetic,Voogt1999MBE,celotto2003characterization,moussy2013epitaxial,liu2014verwey,schemme2014magnetic},
sputtering~\cite{yanagihara2013selective,prieto2015role,xiang2010epitaxial,prieto2016fourfold}
and pulsed laser deposition
(PLD)~\cite{arenholz2006anisotropic,hamie2012investigation,sanz2013stoichiometric,monti2013room,dho2016substrate},
and also on different substrates such as MgO, Al$_2$O$_3$,
MgAl$_2$O$_4$, BaTiO$_3$ and SrTiO$_3$. However, there is at
present no general consensus regarding the magnetic behavior of
magnetite films. An open question is to what extent the
preparation of Fe$_3$O$_4$ films can affect their detailed
magnetic properties, including remanence, coercive field, and
magnetic anisotropy symmetry.

In general, device applications based on magnetic nanostructures
require both understanding and control of the magnetic behavior of
artificially grown films, where the magnetic properties can differ
from the bulk ones and, in addition, can be influenced by the film
growth microstructure, including interfacial strain, crystal
orientation and thickness. The magnetization bulk easy-axis
directions of Fe$_3$O$_4$ at room temperature (RT) are the cubic
$\langle111\rangle$ ones (dashed red vectors in
Fig.~\ref{fig:bulkMagnetite}.b). Thus, in the (001) surface of
bulk samples, the magnetization is expected to lie along the
in-plane $\langle110\rangle$ directions (blue vectors in
Fig.~\ref{fig:bulkMagnetite}.b), i.e., the projection of the bulk
$\langle111\rangle$ on the (001) surface, as confirmed by
magnetic microscopy observations~\cite{FigueraBulkmagnetite}. In
the case of artificially grown magnetite films, many works have
focused on the study of in-plane magnetic anisotropy induced by
the
substrate~\cite{van1998ferromagnetic,horng2004magnetic,nagahama2014magnetic,chichvarina2015magnetic,dho2015distinctive,cheng2008magnetic,kale2001film,hassan2009epitaxial,zhang2011plane,bruns2013fe3o4,monti2013room,xiang2010epitaxial}.
The MgO substrates have been widely
explored~\cite{van1998ferromagnetic,Voogt1999MBE,horng2004magnetic,nagahama2014magnetic,chichvarina2015magnetic,dho2015distinctive,cheng2008magnetic,kale2001film}
due to its low lattice mismatch with Fe$_3$O$_4$ ($+0.3\%$). In
general, in-plane magnetic anisotropies have been reported for
magnetite grown on MgO(001) and MgO(110)
substrates~\cite{van1998ferromagnetic,horng2004magnetic,nagahama2014magnetic,chichvarina2015magnetic,dho2015distinctive,cheng2008magnetic,kale2001film}.
However, in-plane isotropic behavior has been also
reported~\cite{cheng2008magnetic}. Recently,
some works reported perpendicular magnetic anisotropy in
Fe$_3$O$_4$ films deposited on MgO(111)
substrates~\cite{chichvarina2015magnetic}. Other oxide
(001)-oriented substrates with larger lattice mismatch have been
used, such as MgAl$_2$O$_4$ (misfit
$-3.9\%$)~\cite{dho2016substrate}, SrTiO$_3$ (misfit
$-7.0\%$)~\cite{dho2015distinctive,prieto2015role,cheng2008magnetic,kale2001film},
and LaAlO$_3$ (misfit $-9.7\%$)~\cite{prieto2015role}, resulting in a
well-defined magnetic anisotropy of very small strained
Fe$_3$O$_4$ films ($<1\%$) with thicknesses above 40~nm. In
addition, semiconductor single crystals substrates such as
GaAs(001)~\cite{zhang2011plane},
InAs(001)~\cite{huang2012magnetic}, and buffered Si(001)~\cite{hassan2009epitaxial,prieto2016fourfold}, as well as
metallic substrates~\cite{bruns2013fe3o4}, have been also used to
induce in-plane anisotropy.

\begin{figure}[bp]
    \begin{center}
        \includegraphics[width=0.95\textwidth]{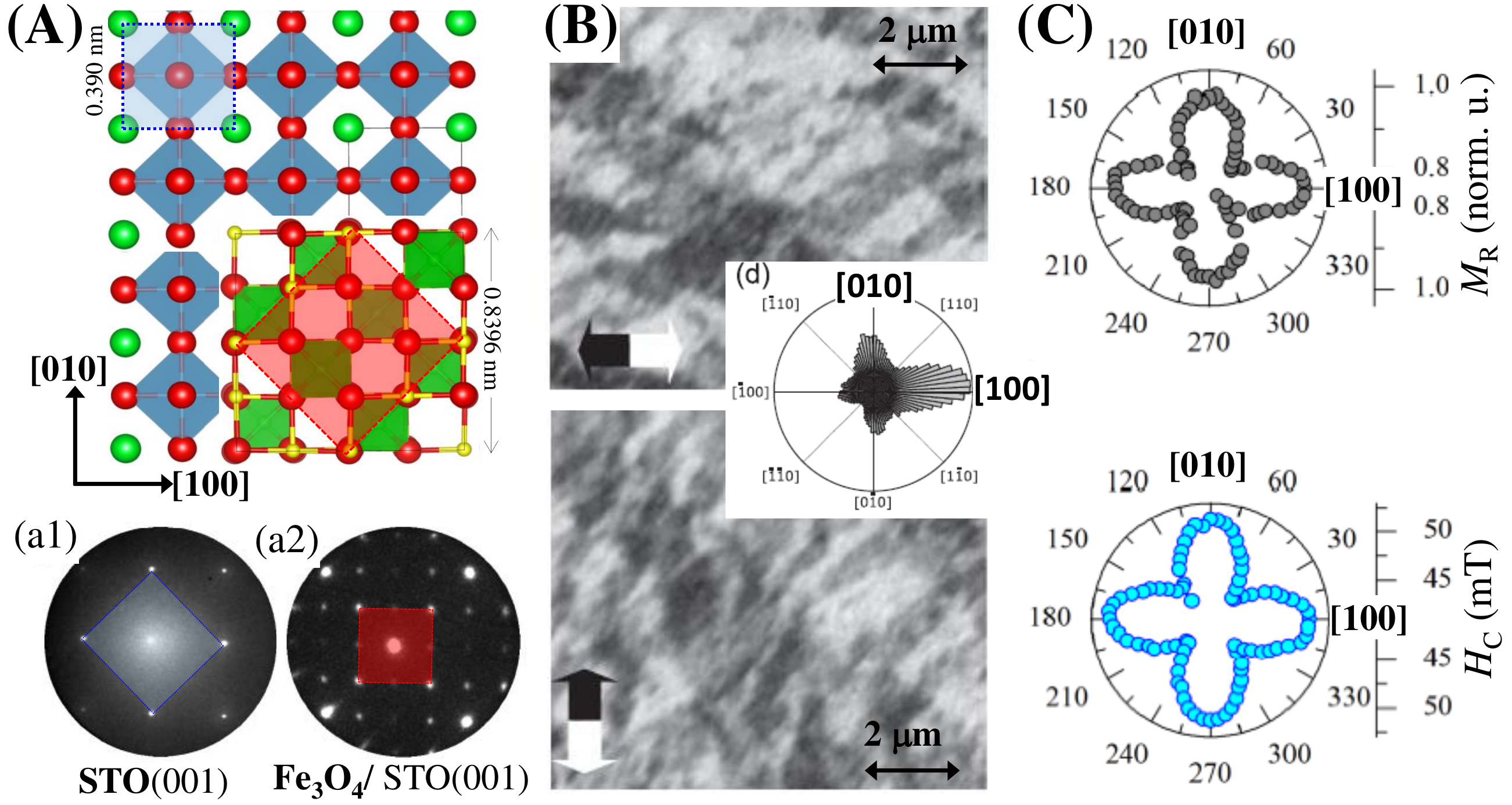}
    \end{center}

    \caption{Identifying magnetic easy-axis directions of stoichiometric magnetite films grown on
    SrTiO$_3$:Nb(001)~\cite{monti2013room}. (A) Epitaxial
    relationship of magnetite on SrTiO$_3$. Oxygen atoms are shown
    as red spheres, with Sr atoms represented by green ones (Ti
    atoms are below in the middle of the blue-grey octahedral).
    The magnetite unit cell is shown in the lower-right side, with
    octahedral irons shown in yellow, and tetrahedral irons shown
    as green filled tetrahedra (schematics prepared by VESTA~\cite{VESTA}).
    The surface unit cells of both materials are drawn by blue and
    red squares, respectively (a1) and (a2) Corresponding LEED
    patterns measured by LEEM (both images are at the same scale).
    (B) Room temperature SPLEEM images acquired at the same
    location with the electron spin-direction along the x-axis
    ([100] direction, top image) and y-axis ([0$\bar{1}$0],
    bottom), showing the local surface magnetization component
    along the given direction. The inset shows the corresponding
    polar histogram of the in-plane magnetization as derived from
    the images. (C) Polar plot representation of the remanence
    (top graph) and coercivity (bottom) derived from kerr
    hysteresis loops acquired at different applied field angles
    with respect to the [100] crystal direction (Adapted from~\cite{monti2013room}).}
    \label{fig:montipaper}
\end{figure}

However, there is no consensus about the easy axis direction in
the films showing what in-plane anisotropic behavior is. In most
cases, the $\langle110\rangle$ surface in-plane directions are
reported for the easy-axis, as expected from a (001)-oriented
surface of bulk magnetite (see Fig.~\ref{fig:bulkMagnetite}.b),
but they do not identify the local domain magnetization direction.
Recently, we showed $\langle100\rangle$ magnetization easy axes in
magnetite films grown on SrTiO$_3$:Nb(100)~\cite{monti2013room}.
The main result of this work is shown in
Fig.~\ref{fig:montipaper}. In particular, stoichiometric epitaxial
Fe$_3$O$_4$ thin films grown with PLD on SrTiO$_3$:Nb(001)
substrates (Fig.~\ref{fig:montipaper}.A) showed individual domains
with magnetization lying mostly along the in-plane
$\langle100\rangle$ directions (Fig.~\ref{fig:montipaper}.B),
while the domain walls were aligned with the $\langle110\rangle$
directions. Furthermore, the remanence and the coercivity display an in-plane fourfold
symmetry with the maxima of both along the $\langle100\rangle$ directions
(Fig.~\ref{fig:montipaper}.C). Similar results have been
reported in magnetite films grown by
sputtering~\cite{prieto2015role,prieto2016fourfold} and
PLD~\cite{arenholz2006anisotropic,dho2016substrate} on different
substrates, in which hysteresis loops acquired along different
directions are compared.

Other important magnetic parameters of magnetite films like the
magnetization reversal mechanism and the coercivity are much less
investigated. The magnetization reversal understanding requires either
magnetic imaging and/or a vectorial-analysis of the magnetization
reversal. In turn, the coercivity is not a direct measure of the
strength of anisotropy. It is modified by defects and may
present strong dynamical effects~\cite{cunado2017emergence}, but
its angular dependence can provide information on the dynamic
effective magnetic anisotropy symmetry, and its evolution with
thickness can indicate general trends, if any. Apart of the
angular dependence measurements for specific thicknesses and
substrates mentioned below, we are not aware of the aforementioned
studies that are required, nor general trends have been identified
yet.

From the above discussion, it is well established that the
properties of magnetite films are determined by the film
microstructure, including interfacial strain, crystal orientation
and thickness. Thus, both magnetic and transport properties
are strongly influenced by the synthesis method and growth
conditions. For instance, poor quality films present the Verwey
transition at a lower
temperature~\cite{aragon1985influence,liu2014verwey}. Moreover,
several results indicate that the magnetism and magneto-transport
phenomena in magnetite films are controlled by antiphase domain boundaries
(APBs) generated during the film growth~\cite{margulies1997origin,celotto2003characterization}.
In fact, the crystal domain size or the APB density seems to be
dependent on the thickness~\cite{eerenstein2002spin,kale2001film}
and on the misfit with the substrate~\cite{cheng2008magnetic}. Therefore, it is clear that
precise control over the microstructure and phase purity is
necessary to control the magnetic behavior, and therefore the
transport behavior. Furthermore, in order to disentangle the magnetic
symmetry orientation and to study its dependence with the
aforementioned microstructure parameters, a study of the
angular evolution of the magnetic properties is required.

In this work, we study pure magnetite films grown by IR-PLD upon changing substrate, orientation, and thickness. The
morphology, microstructure, stoichiometry, Verwey transition, and
angular-dependent magnetic behavior of all films prepared have
been correlated in order to identify the key parameters
controlling the magnetic properties of high quality nanostructured
magnetite films grown by IR-PLD.

\section{Experimental}
\label{sec:Experimental} High quality magnetite films grown on
different substrates, and with different thicknesses, have been
investigated in detail in order to get the key parameters
controlling the magnetic behavior. Pulsed laser deposition (PLD)
has been used for the deposition~\cite{ashfold2004pulsed}. Both laser
irradiation wavelength and substrate temperature crucially
affect the composition, crystallinity, surface structure and the
magnetic properties of the grown samples. PLD investigations of
magnetite thin films are routinely performed with ultraviolet (UV)\cite{bollero2005influence}
lasers. However we have recently demonstrated the preparation of high quality
magnetite films by using 1064~nm infrared-wavelength irradiation
(IR-PLD)~\cite{sanz2013stoichiometric} from
hematite targets.

For the present study, magnetite films of different thickness,
ranging from 30 to 450~nm, have been prepared on different
surfaces, including substrates with cubic, hexagonal, and
amorphous structure. In order to study the possible influence of
the substrate lattice parameter on microstructure and magnetic
properties of the magnetite films (bulk lattice parameter of
0.8395~nm), we have chosen different (001)-oriented substrates
with different lattice mismatch with magnetite: MgO (mismatch
with magnetite is -0.3~\%, tensile), MgAl$_2$O$_4$ (MAO, +3.9~\%,
compressive), SrTiO$_3$ (STO, +7.3~\%, compressive), and
SrTiO$_3$:Nb (STO:Nb, +7.3~\%, compressive). As examples of non-cubic substrates
we have used amorphous substrate Si/SiO$_x$ and hexagonal
Al$_2$O$_3$(0001) substrates.

The Q-switched Nd:YAG laser had a full width at
half-maximum of 15~ns with a 10~Hz repetition rate at a typical
fluence of 4~J/cm$^2$. The substrates were heated to 750~K during
deposition. Layer thicknesses were determined ex situ either by
X-Ray fluorescence and/or by X-ray reflectometry. The morphology,
microstructure, and stoichiometry of the films were characterized
at room temperature (RT) by atomic force microscopy (AFM), X-Ray
diffraction (XRD) and M\"{o}ssbauer spectroscopy, respectively.
M\"{o}ssbauer data were recorded in the electron detection mode
using a conventional constant acceleration spectrometer equipped
with a $^{57}$Co(Rh) source and a parallel plate avalanche counter
~\cite{CEMS}. The velocity scale was calibrated using an
$\alpha$-iron foil and the isomer shifts were referred to the
centroid of the centroid of the RT spectrum of $\alpha$-iron.

Magnetic characterization was carried out by superconducting
quantum interference device magnetometry (SQUID) and
vectorial-resolved magneto-optical Kerr effect (v-MOKE). SQUID
magnetometry is employed to measure the evolution of magnetization
with temperature, which has been used to determine the Verwey
transition of the investigated magnetite thin films. The angular
dependence of the Fe$_3$O$_4$ films magnetic properties was
measured at room temperature (RT) by high-resolution v-MOKE
measurements in a longitudinal
configuration~\cite{jimenez2014vectorial,cunado2015note}. The
hysteresis loops were recorded by changing the in-plane angular
rotation of the sample ($\alpha_H$) and keeping fixed the external
magnetic field direction, which is applied parallel to the film
plane. The whole angular range was probed at intervals of
4.5$^{\circ}$, i.e., from $\alpha_H=0^{\circ}$ to 360$^{\circ}$.

\section{Results}

\begin{figure}[bp]
    \begin{center}
        \includegraphics[width=0.95\textwidth]{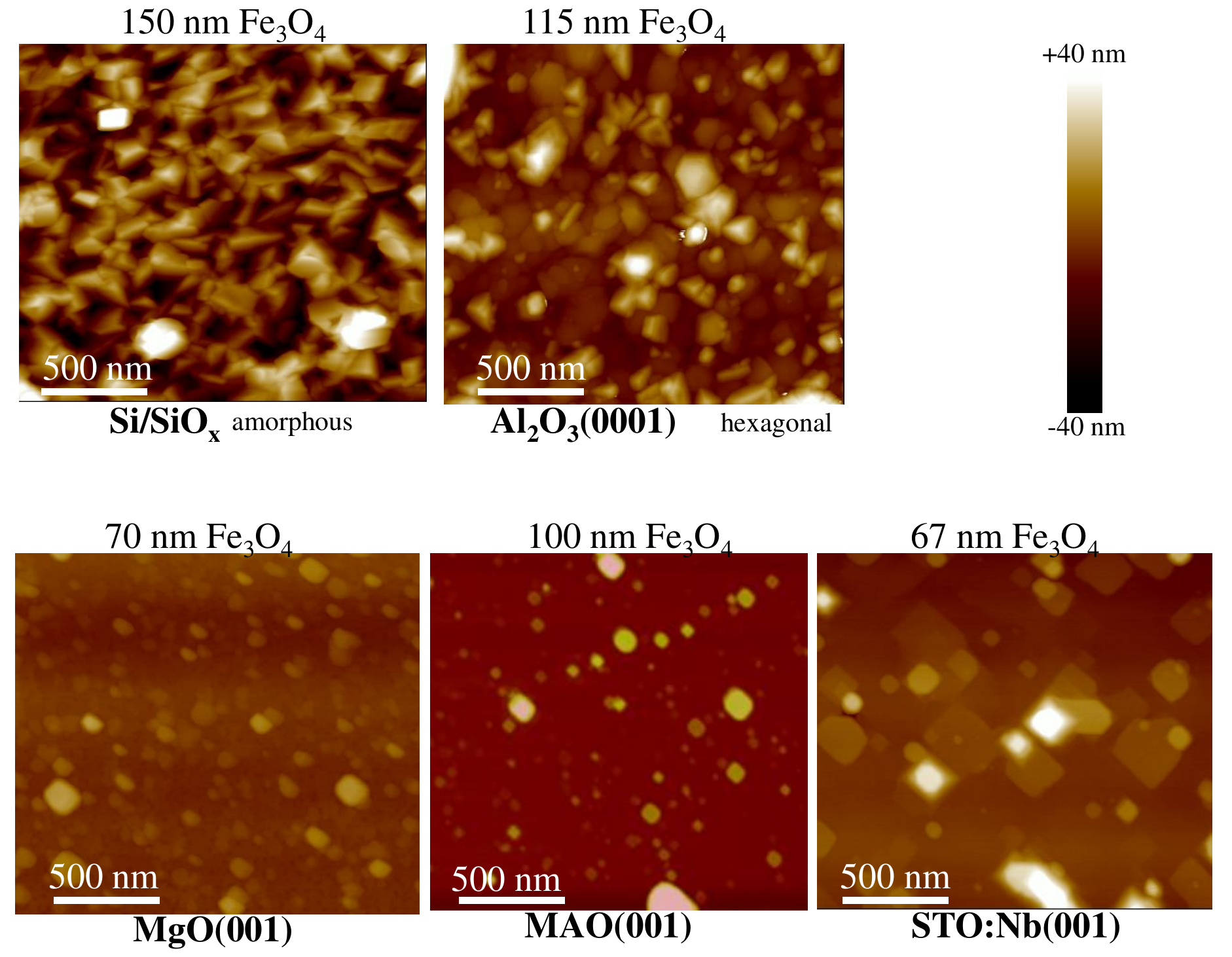}
    \end{center}
    \caption{Surface topography characterization. Selected AFM
    topographic images, $2.0\mu$m wide of Fe$_3$O$_4$ thin films
    grown on non-cubic (top images: amorphous SiO$_2$ and
    hexagonal Al$_2$O$_3$(0001)) and cubic (bottom: MgO(001),
    MgAl$_2$O$_3$(001) and SrTiO$_3$(001)) substrates. The thermal
    color height scale is displayed in the right top side and
    corresponds to 40~nm.} \label{fig:AFM}
\end{figure}

\begin{figure}[tp]
    \begin{center}
        \includegraphics[width=0.95\textwidth]{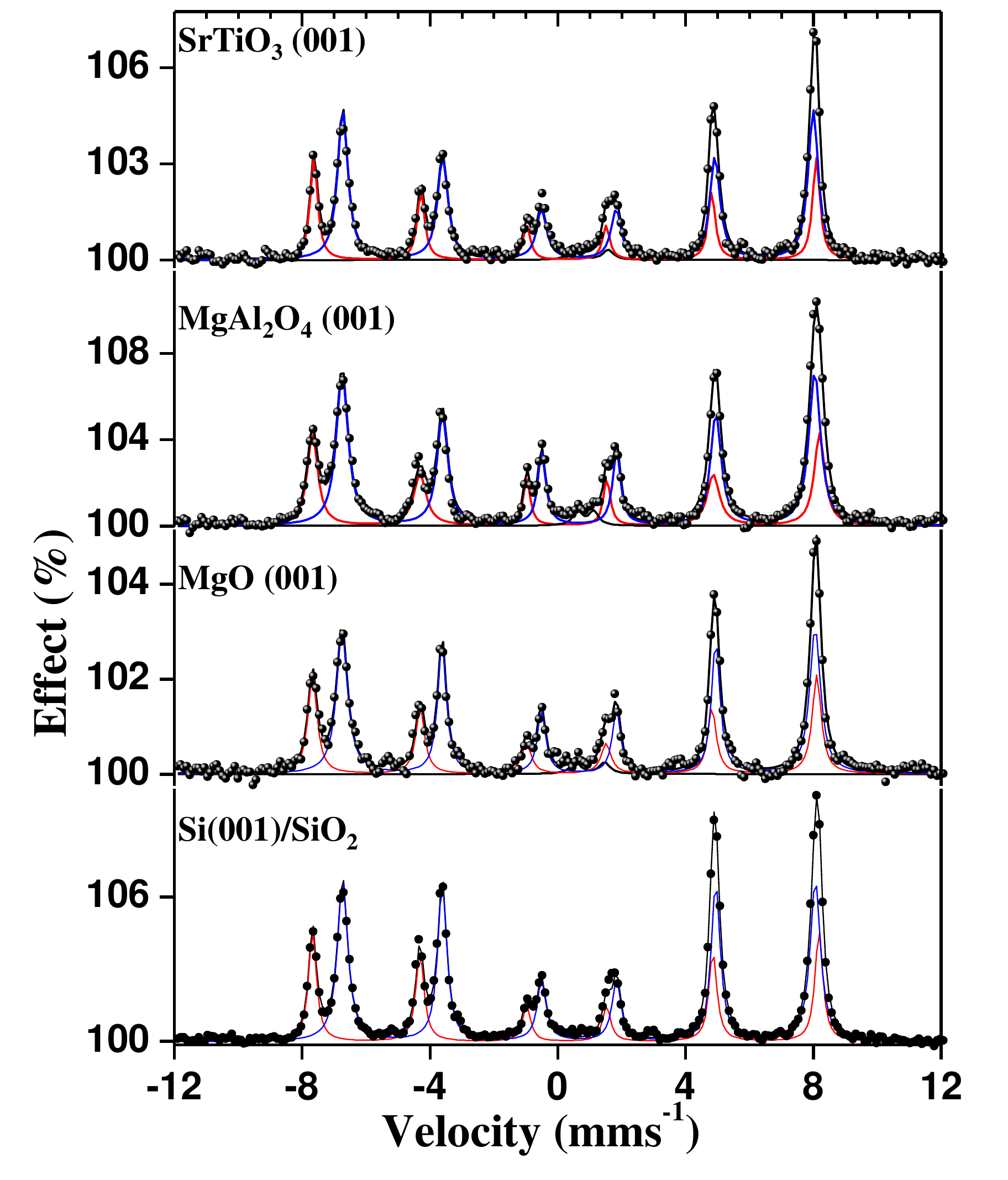}
    \end{center}
    \caption{Room temperature ICEMS spectra recorded from the Fe$_3$O$_4$ films grown on the indicated surfaces.
   The symbols are the experimental data acquired in the films used in Fig.~\ref{fig:AFM}.
   Continuous black lines are the best fits with the two sextet components, as expected for
   magnetite. The corresponding resonances $S_{\rm A}$ and $S_{\rm B}$ are depicted with solid
   red and blue lines, respectively. Note that the ratios $S_{\rm B}/S_{\rm A}$ found indicate that the
   films have stoichiometric composition while the area ratio of the lines 2 and 3 of
   sextets indicate the existence of out-of-plane magnetization components.
    } \label{fig:ICFMS}
\end{figure}

In the following, relevant results on the morphology, chemical,
and structural characterization of the different films
investigated will be presented. Representative microscopy AFM
images of the magnetite films grown on different substrates are
shown in Fig.~\ref{fig:AFM}.  In general, the root-mean-square
roughness (roughness in the following) is much larger in the case
of non-cubic substrates ($>20$~nm), where pronounced elongated,
squared, and orthorhombic features for films hundred nanometer
thick with sharp edges can be identified (first two left images of
Fig.3). Smaller features are found in the film grown on the
Al$_2$O$_3$(0001) surface. For the films deposited on
(100)-oriented substrates, squared shape features ("mesas"), with
heights of up to 30~nm and lateral sizes in the 100-200~nm range,
emerge from a very flat film (with roughness $<3$~nm), similar to
our previous results on STO:Nb~\cite{monti2013room}. Moreover, the
roughness increases with the lattice mismatch and with the
thickness, suggesting that the lattice mismatch at the interface
between both lattices is the main factor controlling the final
morphology.

The good stoichiometry of the films has been checked by integral
conversion electron M\"{o}ssbauer spectroscopy (ICEMS)
measurements at RT. The ICEMS spectrum of stoichiometric magnetite
is composed by two sextets corresponding to the iron
ions located in the octahedral ($S_{\rm B}$) and tetrahedral
($S_{\rm A}$) positions, respectively, of the spinel-related
structure~\cite{vandenberghe2000mossbauer}. The ICEMS spectra
recorded from various representative PLD films are depicted in
Fig.~\ref{fig:ICFMS} and all show these two sextets. For the sake
of consistency all the spectra were fitted with the same criteria,
basically maintaining the linewidth to be equal for the six line
of sextets (but different for the $S_{\rm A}$ and $S_{\rm B}$
components) and the areas in the ratio 3:$x$:1:1:$x$:3, allowing
$x$ to refine until obtaining the best $\chi^2$ value. In general,
the M\"{o}ssbauer parameters obtained from the fit of the spectra
are totally characteristic of magnetite with $x$ values around 2,
between 1.8 (MgAl2O4) and 2.6 (Si) (being 2.2 for both MgO and
STO). The latter is indicative that the films have out-of-plane
magnetization components, where $x=4$ corresponds to full in-plane
magnetization and $x=0$ to full out-of plane
magnetization~\cite{juan2019}. The derived
M\"{o}ssbauer parameters are: isomer shifts
($\delta A=0.26$~mms$^{-1}$ and $\delta B=0.65$~mms$^{-1}$)
quadrupole shifts ($2\epsilon_{\rm A}=-0.02$~mms$^{-1}$ and
$2\epsilon_{\rm B}=-0.02$~mms$^{-1}$) and hyperfine magnetic
fields ($H_{\rm A}=49.0$~T and $H_{\rm B}=46.4$~T). The area ratio
$S_{\rm B}/S_{\rm A}$ of the two components is $1.9$ in all the
cases, what indicates that the films are of stoichiometric
composition~\cite{vandenberghe2000mossbauer}. Thus, our data
suggest that the substrate has no influence on the chemical
composition of the films. The different $x$ values found could
be ascribed to the use of different substrates and thicknesses, as
discussed in the next sections.

%\newpage

\subsection{Influence of the substrate}
\label{subsec:Influence_substrate} The structural and magnetic
characterization of magnetite films with similar thickness grown
on different substrates are correlated and compared, in order to
study the effect of the film's microstructure on its magnetic
properties.

\begin{figure}[tp]
    \begin{center}
        \includegraphics[width=0.8\textwidth]{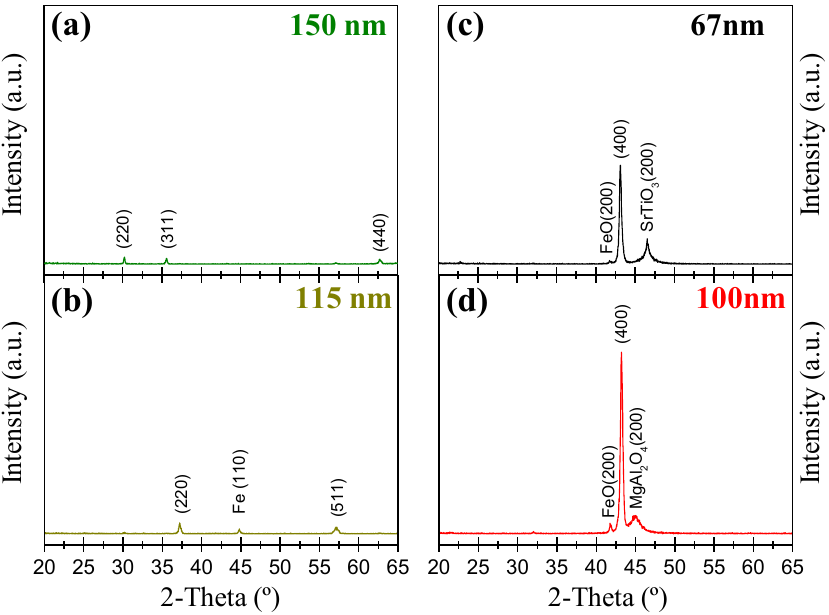}
        \caption{Selected XRD patterns of Fe$_3$O$_4$ thin films
        deposited on (a) Si(001) with a native SiO$_2$ layer, (b)
        Al$_2$O$_3$(0001), (c) SrTiO$_3$:Nb(001) and (d)
        MgAl$_2$O$_4$(001). } \label{fig:XRD}
    \end{center}
\end{figure}

Figure~\ref{fig:XRD} shows representative XRD patterns of
Fe$_3$O$_4$ films grown on non-cubic substrates (a and b) and on
(001)-oriented cubic substrates (c and d). The films
grown on non-cubic substrates, i.e., on amorphous SiO$_2$ and
Al$_2$O$_3$(0001), display peaks located at
30.2$^{\circ}$, 35.5$^{\circ}$, 37.1$^{\circ}$,
57.2$^{\circ}$ and 62.7$^{\circ}$, which are assigned to
Fe$_3$O$_4$ reflections (220), (311), (220), (422), (511) and
(440) planes, respectively (Fig.~\ref{fig:XRD}.a and .b). This
pattern is indicative of polycrystalline magnetite (Joint
Committee on Powder Diffraction Standards Card No. 88-0866). An
additional peak at 44.67$^{\circ}$ is observed
on the film grown on Al$_2$O$_3$ substrate
(Fig.~\ref{fig:XRD}.b). This peak is assigned to metallic iron.
In contrast, the films grown on
(001)-oriented cubic substrates (SrTiO$_3$:Nb and MgAl$_2$O$_4$)
show a well-defined preferential orientation related to the
surface of the substrate, as can be observed by the predominance
of (400) reflection peak at an angle 43.3$^{\circ}$
(Fig.~\ref{fig:XRD}c and ~\ref{fig:XRD}d). In this case, the films
exhibit a small contribution of W\"{u}stite (Fe$_{0.8}$O$_{0.2}$)
that can be observed at an angle 41.8$^{\circ}$. By
comparing the intensity of peaks obtained for both families of
substrates it can be noticed the high quality (crystallinity) of
Fe$_3$O$_4$ films obtained when a (001)-oriented cubic substrate
was employed.

Several structural parameters have been derived from the XRD
patterns of Fig.~\ref{fig:XRD}. The average crystallite size have
been calculated for all samples using Scherrer formula, resulting
the following sizes: 28, 23, 31 and 50~nm for the
Fe$_3$O$_4$ films grown on SrTiO$_3$:Nb, MgAl$_2$O$_4$,
Al$_2$O$_3$ and SiO$_2$ substrates, respectively. The
polycrystalline films have larger crystallite sizes when compared
with the epitaxial ones of similar thickness. Moreover, larger
crystallite sizes are found for the case of STO when compared with
the other (001)-oriented cubic substrates. XRD patterns have been
used as well to calculate the lattice parameter of the magnetite
films, based on Bragg's law, obtaining values of 0.836, 0.836,
0.836 and 0.833~nm for SiO$_2$, Al$_2$O$_3$, MgAl$_2$O$_4$ and
SrTiO$_3$:Nb substrates, respectively. Notice that all these
values are quite close to the bulk one(0.840~nm), but with a
small compressive strain ($<$ 1$\%$).

\begin{figure}[tp]
    \begin{center}
        \includegraphics[width=0.8\textwidth]{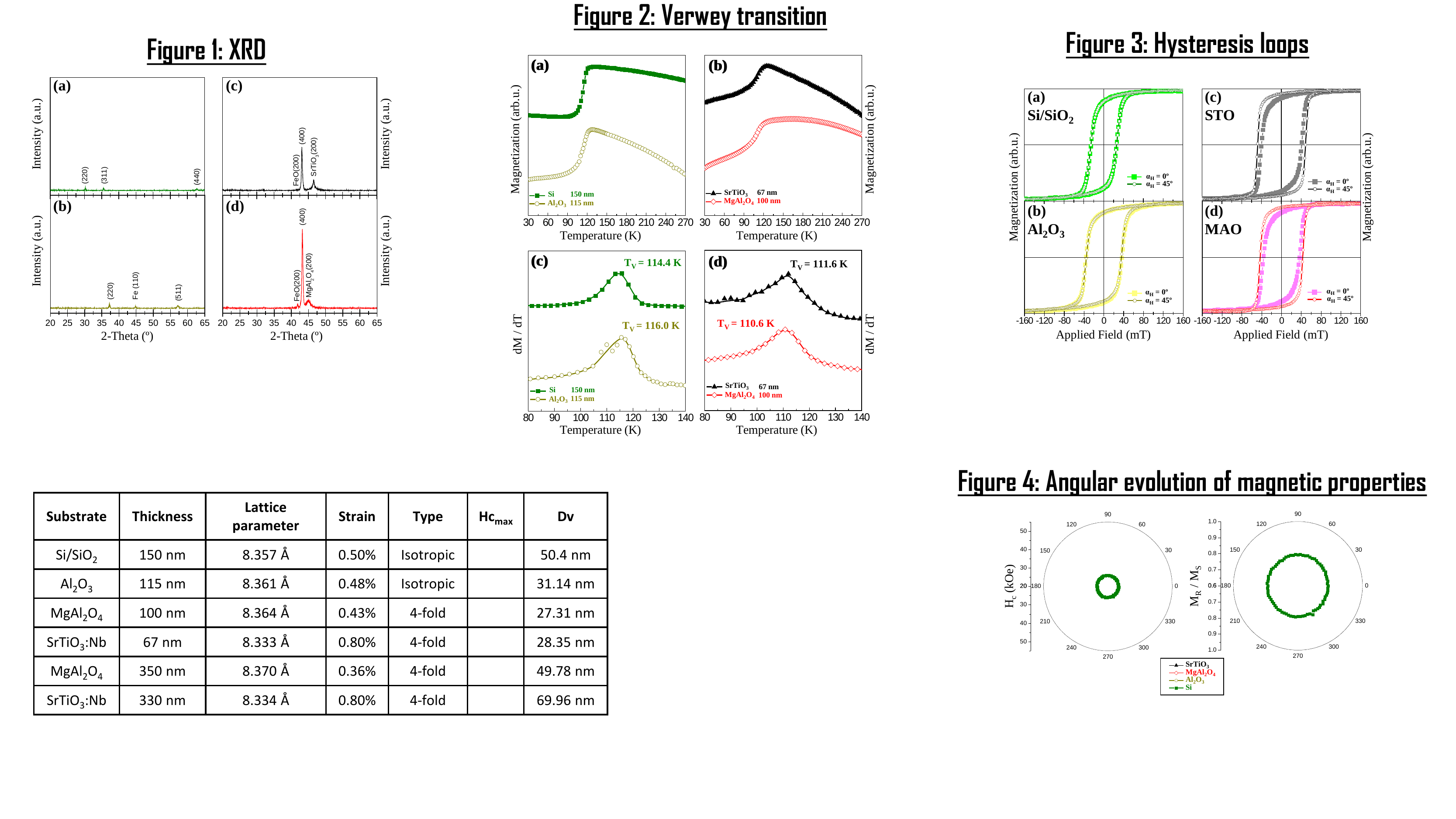}
        \caption{Determination of the Verwey transition
        temperature $T_{\rm v}$ for the polycrystalline (left
        graphs) and epitaxial (right) Fe$_3$O$_4$ films used in
        Fig.~\ref{fig:XRD}. In the top graphs the
        evolution of the films magnetization during the warming
        with an in-plane applied field of 2~kOe is plotted. On the bottom
        graphs, the corresponding temperature evolution of
        d$M$/d$T$ is plotted. For clarity, different colors and shapes have
        been used for the different Fe$_3$O$_4$ films: Si/SiO$_2$ with green square symbols,
        Al$_2$O$_3$ with dark yellow circles, SrTiO$_3$:Nb
        with black triangles; and MgAl$_2$O$_4$ with red
        diamonds, respectively. $T_{\rm v}$ is derived from the maximum value of
        the d$M$/d$T$($T$) curve.} \label{fig:VerweyCH1}
    \end{center}
\end{figure}

The Verwey transition, which as mentioned depends strongly
on crystal quality, was investigated by measuring the
magnetization as a function of temperature. The transition
is defined as the point with
maximum rate of change in magnetization against temperature. A
representative study for magnetite films of similar thickness
grown on different substrates is displayed in
Fig.~\ref{fig:VerweyCH1}. The evolution of the magnetization
during warming of non-cubic and cubic substrates is compared in
Fig.~\ref{fig:VerweyCH1}.a and Fig.~\ref{fig:VerweyCH1}.b,
respectively. In order to identify clearly $T_v$, the evolution in
temperature of the derivative of the magnetization with respect to
the temperature has been plotted in Fig.~\ref{fig:VerweyCH1}.c and
Fig.~\ref{fig:VerweyCH1}.d, respectively. $T_v$ values of 114
and 116~K are obtained for the films grown on SiO$_2$ and
Al$_2$O$_3$, respectively, while for SrTiO$_3$:Nb and
MgAl$_2$O$_4$ the transition is found at 111 and 112~K,
respectively. The difference of temperatures for the Verwey
transition in different films might be related to strain induced
by the substrate during first stages of growth, as well as to the
crystalline average size~\cite{Liu2014}. For instance,
polycrystalline films (with large crystallite size) have Verwey
transition temperatures very close to the bulk magnetite value
(120 K). By comparison, epitaxial Fe$_3$O$_4$ films (smaller
crystallite size) present lower values. The difference also could
be related with the thickness of the different films, because the
epitaxial films are thinner than polycrystalline ones. In fact,
$T_{\rm V}$ increases with increased thickness, as can be seen in
Fig.~\ref{fig:VerweyCH2}.

\begin{figure}[tp]
    \begin{center}
        \includegraphics[width=0.9\textwidth]{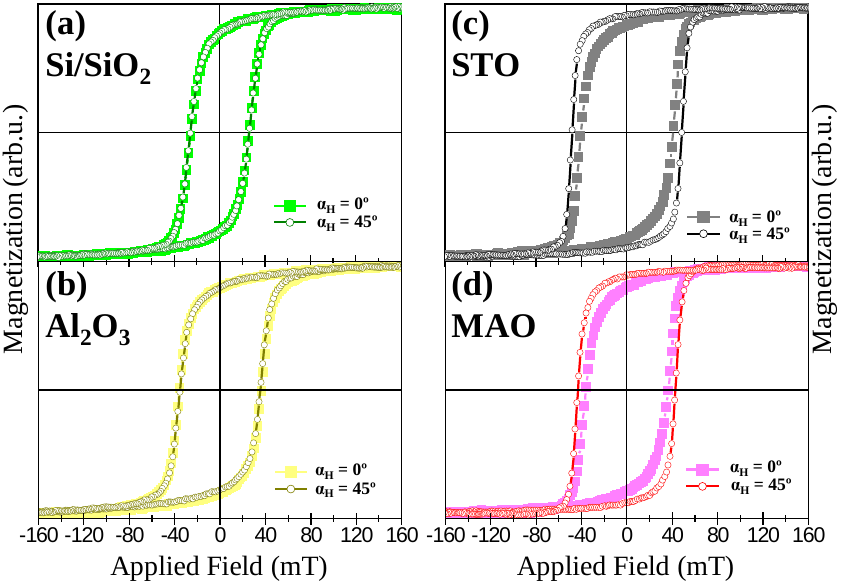}
        \caption {Selected in-plane hysteresis loops of
        Fe$_3$O$_4$ thin films deposited on (a) Si with a native
        SiO$_2$ layer, (b) Al$_2$O$_3$, (c) SrTiO$_3$:Nb and (d)
        MgAl$_2$O$_4$. Graphs (c) and (d) show hysteresis loops
        acquired at $\alpha_H=0^\circ$ and $\alpha_H=45^\circ$,
        which correspond to the easy and hard axis directions,
        respectively.} \label{fig:HystLoops}
    \end{center}
\end{figure}

\begin{figure}[bp]
    \begin{center}
        \includegraphics[width=0.95\textwidth]{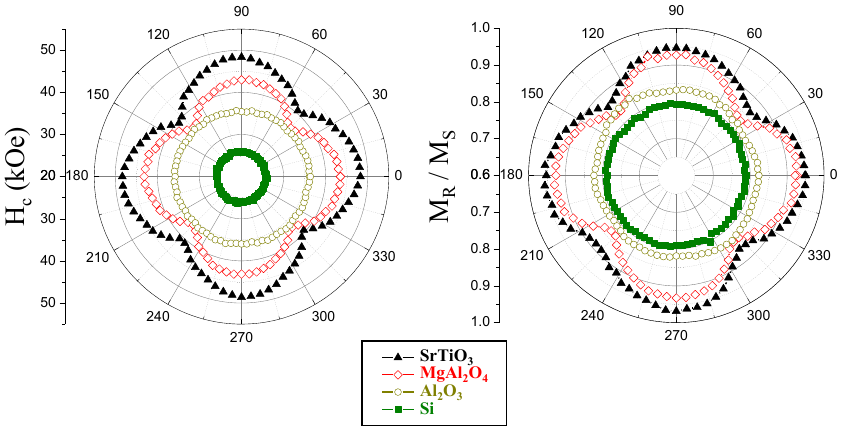}
        \caption{Polar-plot representations of the angular evolution of $H_{\rm C}$
         and $M_{\rm R}/M_{\rm S}$ of Fe$_3$O$_4$ thin films
        deposited on SrTiO$_3$:Nb (full triangles, black),
        MgAl$_2$O$_4$ (open diamonds, red), Al$_2$O$_3$ (open
        circles, dark yellow) and Si with a native SiO$_2$ layer
        (full squares, green). The symbols are derived from the
        corresponding hysteresis loops, as the one shown in
        Fig.~\ref{fig:HystLoops}, acquired at the different
        angular conditions in the whole range. Note the isotropic
        behavior of the polycrystalline magnetite films and the
        well-defined fourfold magnetic symmetry aligned along the
        $\langle100\rangle$ surface directions in the case of the
        epitaxial films.} \label{fig:PolarPlots}
    \end{center}
\end{figure}

Very clear differences can be observed when the magnetic
properties of polycrystalline and epitaxial films are compared.
MOKE hysteresis loops have been systematically recorded by
changing the in-plane orientation of the applied magnetic field
($\alpha_H$) in the whole angular range. Fig.~\ref{fig:HystLoops}
shows representative in-plane hysteresis loops measured at RT for
Fe$_3$O$_4$ films deposited on different substrates. Two loops are
plotted for each sample, with the applied magnetic field aligned
along the $[100]$ ($\alpha_H=0^\circ$) and $[110]$
($\alpha_H=45^\circ$) magnetite surface directions. Both
hysteresis loops are identical for the case of polycrystalline
films, i.e., films grown on SiO$_2$ and Al$_2$O$_3$ substrates,
indicating an isotropic magnetic behavior (see
Fig.~\ref{fig:HystLoops}.a and b). In contrast, a different
magnetic behavior is identified for the two angular conditions in
the case of films grown on SrTiO$_3$:Nb and MgAl$_2$O$_4$
substrates (Fig.~\ref{fig:HystLoops}.c and d). In this case, the
hysteresis loops acquired at the two angular conditions show a
different magnetization reversal pathway, and therefore different
coercivity ($H_{\rm C}$) and remanence ($M_{\rm R}$) values. In
particular, the remanence and coercivity values are higher for
$\alpha_H=0^\circ$, i.e., along the $[100]$ direction. The lowest
coercivity and remanence values are found at $\alpha_H=45^\circ$,
i.e., along the $[110]$ direction. Between these, from
$\alpha_H=0^\circ$ to $45^\circ$, both coercivity and remanence
decrease and the opposite trend is found from $\alpha_H=45^\circ$
to $90^\circ$. In fact, the trend is repeated  every $90^\circ$
for the case of the epitaxial films, as shown in the corresponding
polar-plots representations of coercivity and remanence depicted
in Fig.~\ref{fig:PolarPlots}. These clearly show a fourfold
magnetic symmetry with the maxima along the $\langle100\rangle$
directions. By contrast, the polar-plots for films grown on
SiO$_2$ and Al$_2$O$_3$ display no angular dependence of
coercivity and remanence, indicative of an isotropic magnetic
behavior of the polycrystalline films. This isotropic behaviour is
also reflected (for out-of-plane vs in-plane) in the components of
the M\"{o}ssbauer spectra.

The fourfold symmetry is found for all epitaxial magnetite films
grown on (001)-oriented oxide substrates and oriented along the
$\langle100\rangle$ directions, i.e.,
rotated $45^\circ$ with respect to the
expected $\langle110\rangle$ surface directions (see
Fig.~\ref{fig:bulkMagnetite}). The difference between Fe$_3$O$_4$
films grown on different substrates is the value of coercivity.
For data shown in Fig.~\ref{fig:PolarPlots}, the highest
(smallest) coercivity values are found for the film grown on STO
(Si). The difference in coercivity could be also attributed to the thickness
difference of the films: the thinnest (thicker) film is the one
grown on STO (Si). In the next section it is shown that, in fact,
coercivity depends strongly on the films thickness.

%\newpage

\subsection{Influence of film thickness}
\label{subsec:Influence_thickness}

% FIGURE:
\begin{figure}[bp]
    \begin{center}
        \includegraphics[width=0.85\textwidth]{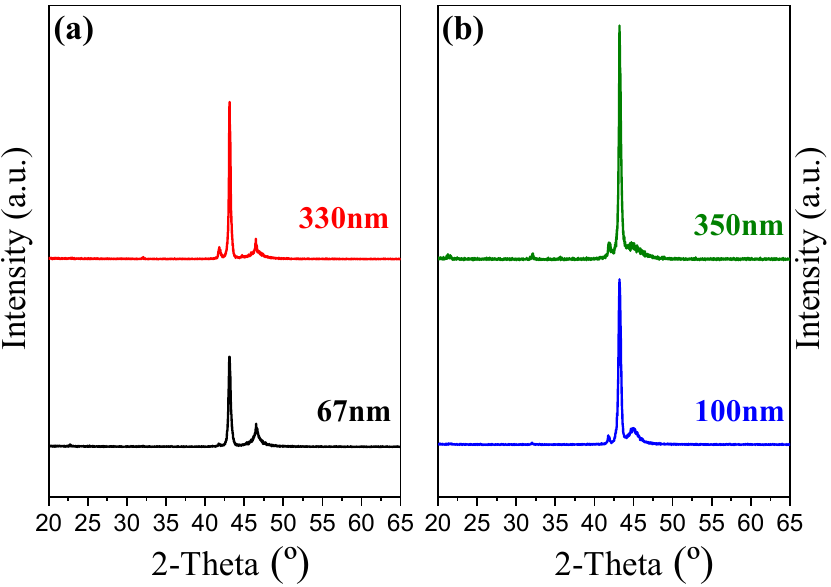}
        \caption{Thickness-dependent XRD diffraction patterns of
        magnetite films grown on SrTiO$_3$:Nb(001) (a) and
        MgAl$_2$O$_4$(001) (b).} \label{fig:XRD_Thickness}
    \end{center}
\end{figure}

Figure~\ref{fig:XRD_Thickness} displays the diffraction patterns
of Fe$_3$O$_4$(001) films of different thickness grown on STO:Nb
and MAO. The increase of the Fe$_3$O$_4$ (400) peak intensity is
clearly related with the increased thickness. Note that even for
thicknesses as large as 330 and 350~nm, the films display a single
crystal orientation, indicating the suitability of the IR-PLD
technique to obtain thick Fe$_3$O$_4$(001) films with high
crystallinity. The calculated strain of the films is similar when
comparing the thin and the thick films grown on the same
substrate. The one corresponding to STO:Nb is a little less
relaxed; the strain on STO:Nb (MAO) is c.a. 0.8$\%$ (0.3$\%$). The
corresponding crystallite sizes, calculated using the Scherrer
equation, are larger for the thicker films. The crystallite sizes
are 28 and 70~nm (27 and 50~nm) for the films grown on STO:Nb
(MAO) with thickness of 67 and 330~nm (100 and 350~nm).

\begin{figure}[tp]
    \begin{center}
        \includegraphics[width=0.85\textwidth]{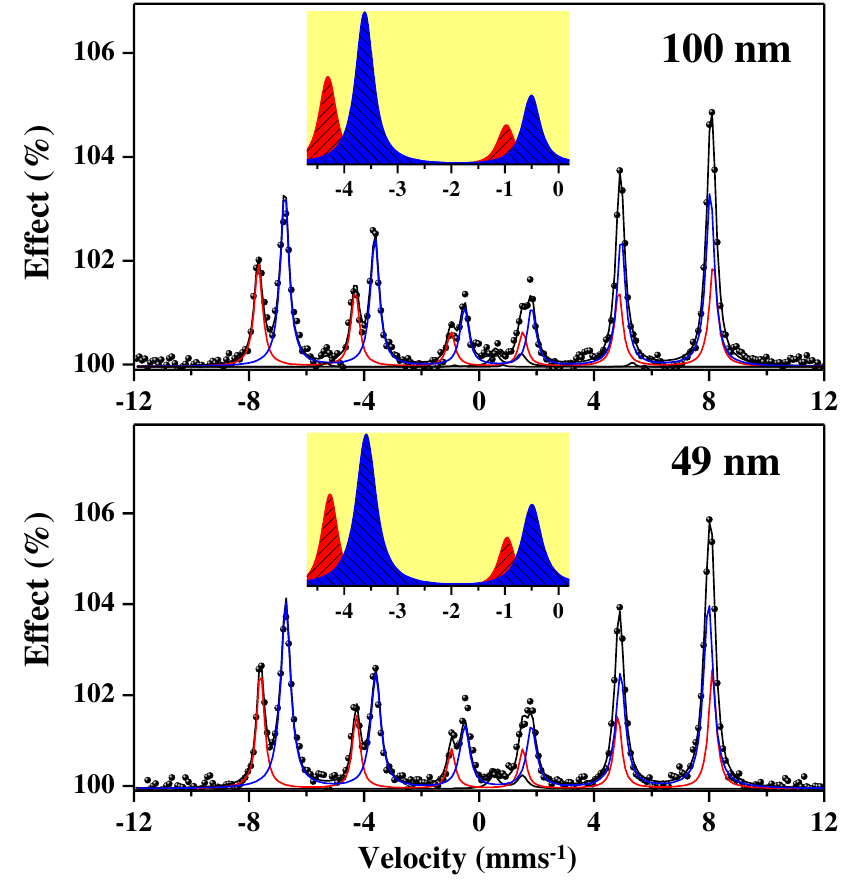}
    \end{center}
    \caption{Thickness-dependent ICEMS spectra of epitaxial Fe$_3$O$_4$ films grown on MgO(001).
   The symbols are the experimental data acquired at RT for the thick (bottom graph) and the thin (bottom) film.
   Continuous black lines are the best fits with the two sextet components, as expected for
   magnetite. The corresponding resonances $S_{\rm A}$ and $S_{\rm B}$ are depicted with solid
   red and blue lines, respectively. $S_{\rm B}/S_{\rm A}=1.9$ for the two films. The insets are zooms of the corresponding
   fits to visualize clearly that the area ratio $x$ of the lines 2 and 3 of sextets
   is lower for the thinner film, suggesting that the magnetization of the thinner film is more out-of-plane.
    } \label{fig:ICFMS_MgO}
\end{figure}

The purity, i.e., good stoichiometry, of the magnetite films have
been checked by ICEMS and by the determination of the Verwey
transition $T_{\rm V}$. In brief, the ICEMS spectra reflect high
stoichiometric magnetite films and there is a non-negligible
variation of $T_{\rm V}$ with both substrate and thickness.
Figure~\ref{fig:ICFMS_MgO} compares ICEMS spectra recorded from
two magnetite films with different thickness deposited on
MgO(001). The spectra were fitted with the same criteria and
M\"{o}ssbauer parameters described before (see
Fig.~\ref{fig:ICFMS}). Similarly, the area ratio $S_{\rm B}/S_{\rm
A}$ found is $1.9$ for both magnetite films, indicating that are
of stoichiometric composition~\cite{vandenberghe2000mossbauer}.
The $x$ parameter, i.e., the area ratio of the lines 2 and 3, is
2.2 for the thicker film and 1.8 for the thinner one, as clearly
show the corresponding insets of Fig.~\ref{fig:ICFMS_MgO},
indicating that the magnetization of the thinner film is more
out-of-plane than in the thicker one.

% FIGURE 6:
\begin{figure}[tp]
    \begin{center}
        \includegraphics[width=0.7\textwidth]{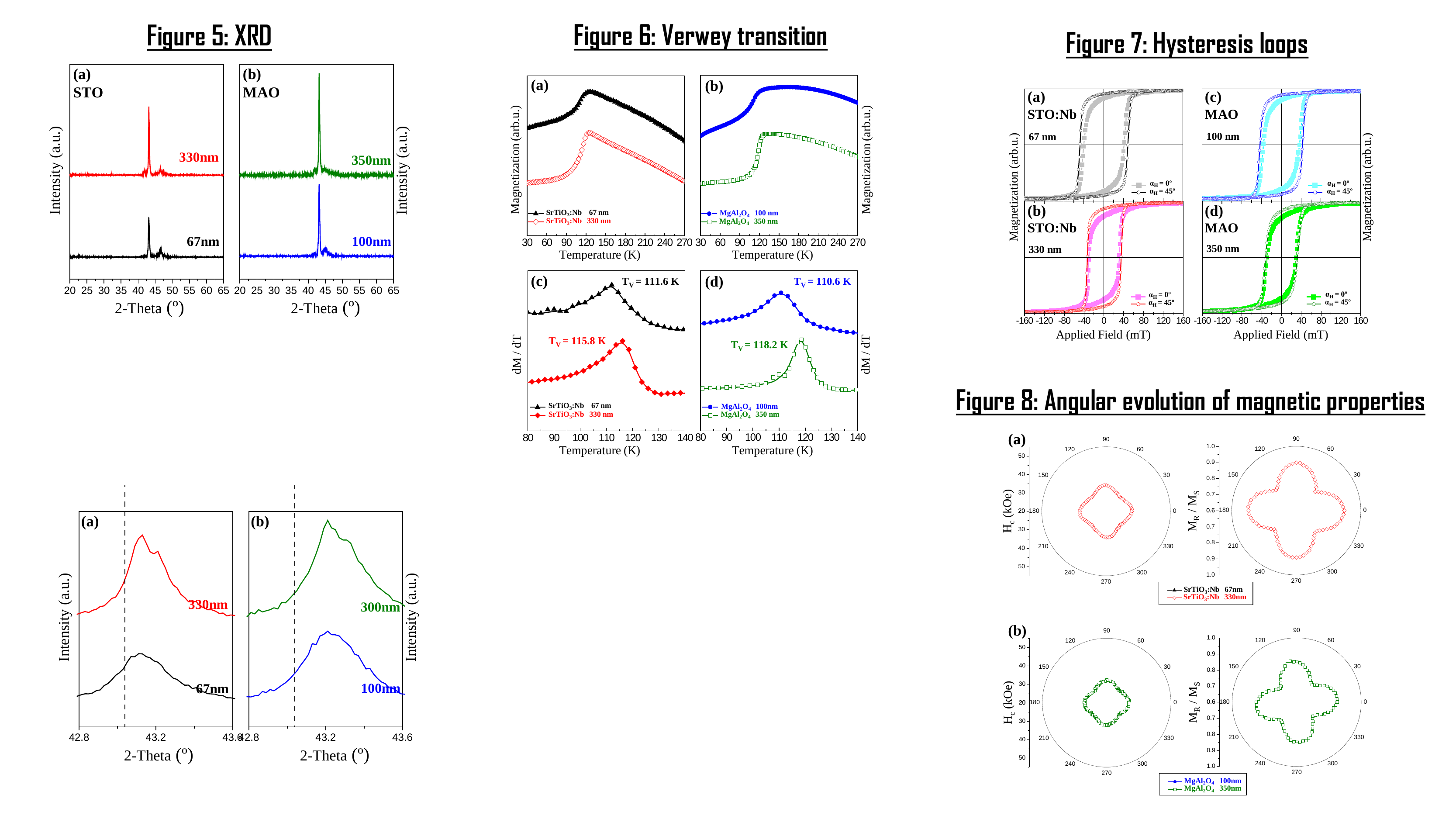}
        \caption{Thickness-dependent Verwey temperature transition
        ($T_{\rm v}$) of films grown on SrTiO$_3$:Nb(001) (left
        graphs, a. and c.) and MgAl$_2$O$_4$(001) (right, b. and
        d.) substrates. In the top graphs are plotted the evolution
        of magnetization of the films during warming with an
        in-plane applied field of 2~kOe. On the bottom graphs are
        plotted the corresponding temperature evolution of d$M$/d$T$.
        For clarity, different colors and shapes have been used
        for the different Fe$_3$O$_4$ films. Note that $T_{\rm v}$
        increases, getting closer to the bulk one, as thickness
        increases.} \label{fig:VerweyCH2}
    \end{center}
\end{figure}

Each of the top graphs of Fig.~\ref{fig:VerweyCH2} compares the
magnetization evolution with temperature for different film
thicknesses grown on different substrates. As it was previously
mentioned, there is a reduction of the Verwey transition
temperature to lower temperatures as the thickness decreases.
Comparing the films grown on STO:Nb and MAO, the latter shows
higher $T_{\rm V}$ values, i.e., closer to the bulk one, probably
due to the smaller strain found on this film. As mentioned in the
introduction, $T_{\rm V}$ is very sensitive to
stoichiometry~\cite{aragon1985influence} but also to
strain~\cite{liu2016fe3o4} and to the presence of structural
defects~\cite{liu2014verwey,dho2016substrate}, even in highly
stoichiometric films. Most work on epitaxial (compressive) films
show a reduced $T_{\rm V}$ (we note that magnetite films prepared
under tensile strain have recently shown substantially higher
values than the bulk~\cite{liu2016fe3o4}). In turn, $T_{\rm V}$
decreases as the number of defects
increases~\cite{eerenstein2003diffusive}. For example, magnetite
films often present antiphase domain boundaries
(APBs)~\cite{margulies1997origin,eerenstein2003diffusive}, as
discussed later. APBs densities can be related to the deposition
temperature~\cite{eerenstein2003diffusive}, and decreased by
post-deposition annealing. So, for films with compressive strain
grown on different substrates, the larger strain (films grown on
STO:Nb), the lower $T_{\rm V}$. In addition, thick films are at
high temperature for longer times during fabrication, which should
result in fewer defects and thus a larger $T_{\rm V}$. The other
observed dependenciy, i.e., of the out-of-plane magnetization
component, will be analyzed when discussing the dependence of
magnetic properties with the film thickness.

% FIGURE 7:
\begin{figure}[tp]
    \begin{center}
        \includegraphics[width=0.9\textwidth]{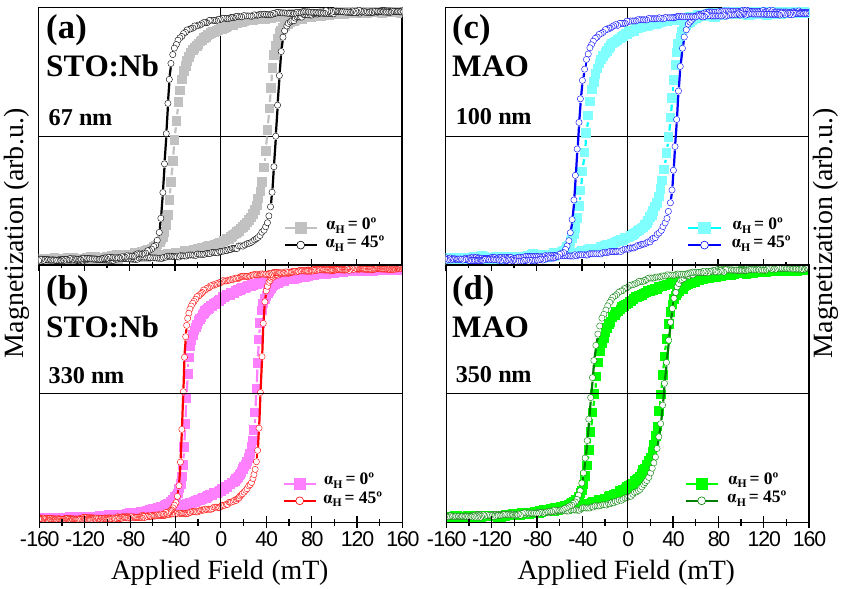}
        \caption{Selected in-plane hysteresis loops (easy and hard axis) comparing the influence of the thickness of epitaxial Fe$_3$O$_4$ films grown on STO:Nb(001) (left graphs) and MAO(001) (right). (a) 67~nm and (b) 330~nm deposited on SrTiO$_3$:Nb, and (c) 100~nm and (d) 350~nm deposited on MgAl$_2$O$_4$.}
        \label{fig:Hyst}
    \end{center}
\end{figure}

\begin{figure}[tp]
    \begin{center}
        \includegraphics[width=0.94\textwidth]{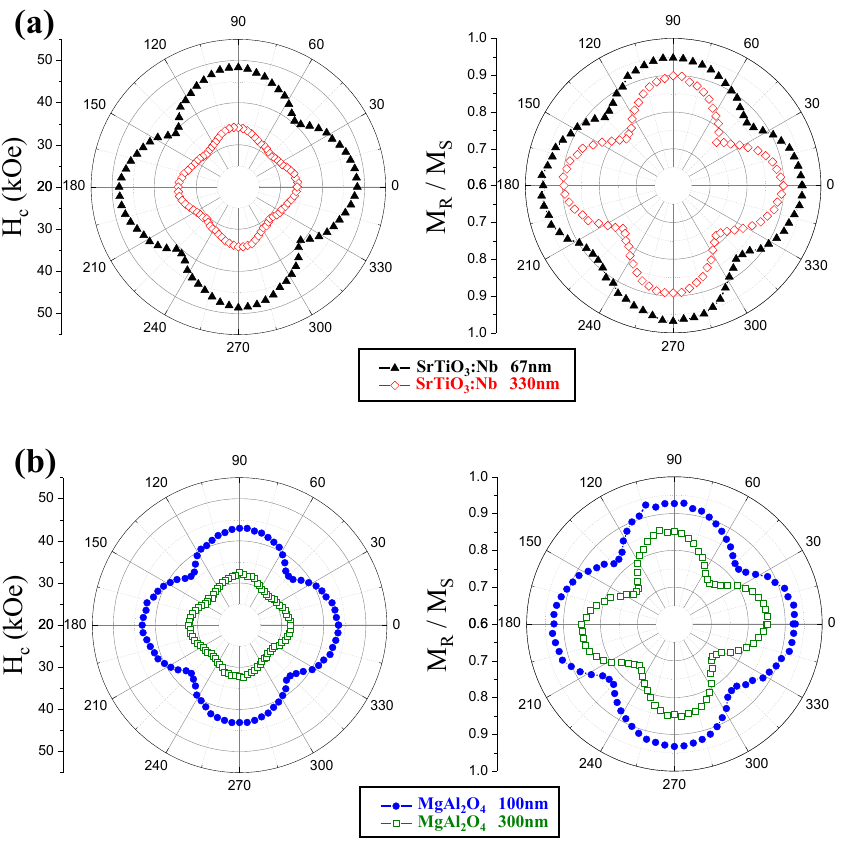}
        \caption{Angular evolution of in-plane \textit{H$_C$} and
        \textit{M$_R$/M$_S$} of Fe$_3$O$_4$ thin films deposited
        on SrTiO$_3$:Nb substrates with thickness of 67~nm (full
        triangles, black) and 330~nm (open diamonds, red), and
        MgAl$_2$O$_4$ substrates with thickness of 100~nm (full
        circles, blue) and 350~nm (open squares, green).}
        \label{fig:Hystpolar2}
    \end{center}
\end{figure}

Representative hysteresis loops acquired at $\alpha_{\rm H}=
0^{\rm \circ}$ and $45^{\rm \circ}$ comparing the magnetic
behavior of Fe$_3$O$_4$ films with different thickness are
displayed in Fig.~\ref{fig:Hyst}.  The left (right) graphs
correspond to films grown on STO:Nb (MAO). The top (bottom) graphs
compare loops of the thinner (thicker) films. All
hysteresis show single loop behavior, and the thicker the film the
smaller the coercivity and the remanence. In addition, the fourfold
symmetry is clearly observed for all films grown on the
(001)-oriented substrates, with the
maxima of both coercivity and remanence lying along the
$\langle100\rangle$ surface crystal directions, and the
corresponding minima along the $\langle110\rangle$ directions.
This magnetic symmetry is clearly observed in the polar plot
representation of both coercivity and remanence shown in
Fig.~\ref{fig:Hystpolar2}. Each polar-plot compares different
thickness. The trend occurrs for the whole angular range.
The coercivity and remanence are smaller in the case of the
thicker films. The fourfold magnetic symmetry is preserved for
magnetite thicknesses as large as 450~nm.

% FIGURE 8:

\begin{figure}[tp]
    \begin{center}
        \includegraphics[width=1.0\textwidth]{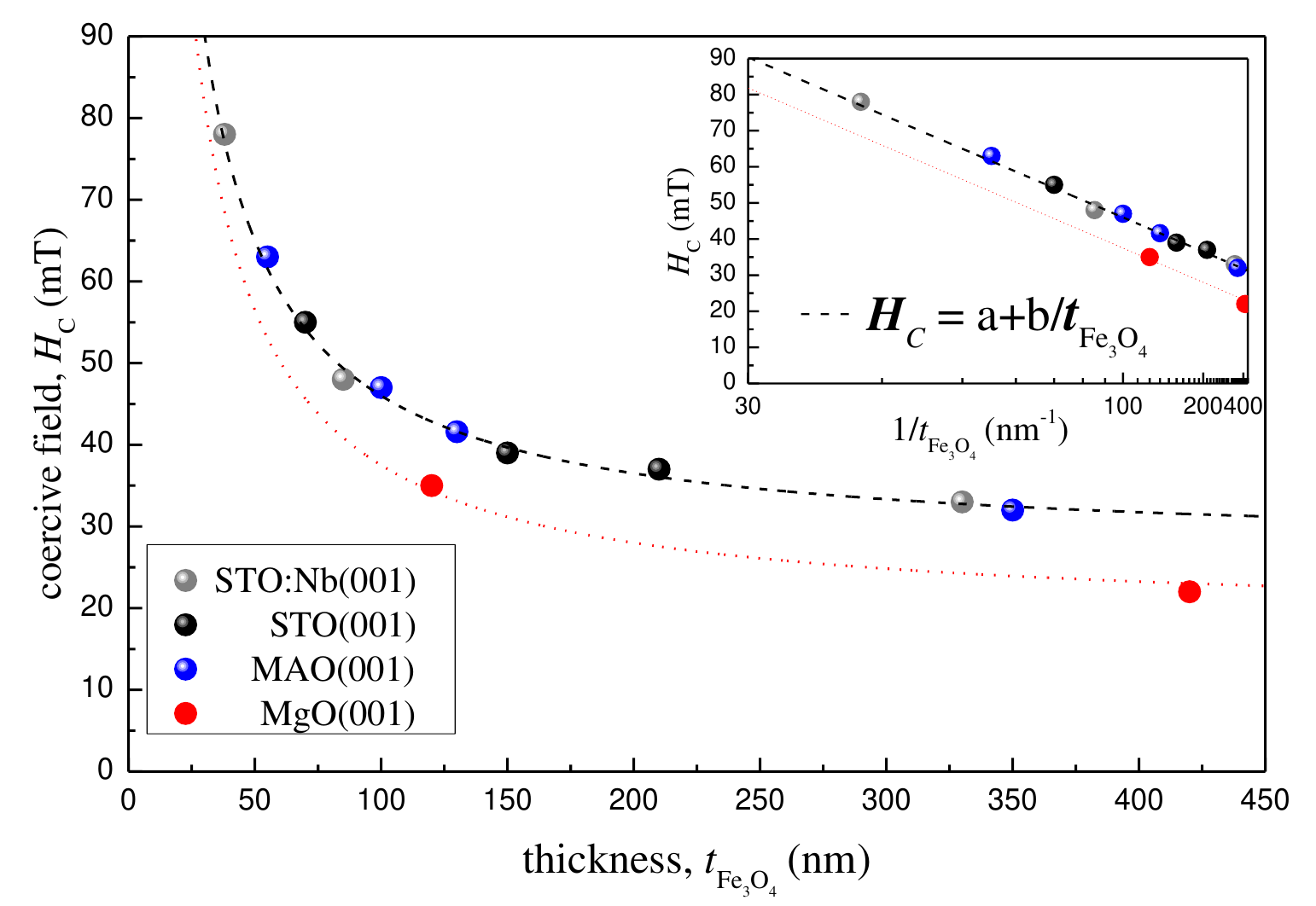}
        \caption{Thickness-dependent coercivity of epitaxial
        magnetite films. Inset: corresponding coercivity $vs.$
        1/thickness plot. The experimental $H_{\rm C}$ values
        (symbols) have been derived from the hysteresis loops
        acquired with the field along the [100] direction, i.e.,
        e.a., as those shown in Fig.~\ref{fig:Hyst}. Different colors have been
        used for the films grown on different substrates. The
        dashed line is a fit to the coercivity using an inverse
        law with thickness, as discussed in the text.}
        \label{fig:Hcthickness}
    \end{center}
\end{figure}

The thickness-dependent coercivity along the easy direction for
all films investigated has been plotted in
Fig.~\ref{fig:Hcthickness}. At a glance, the coercivity
decreases as the film
thickness increases, independently of the substrate used. A
similar scenario but with smaller coercivity values is found for
the other angular conditions. The coercivity follows
an inverse law with the film thickness, whose accuracy is evident in the inset of the figure.

\section{Discussion}

The substrate symmetry is related to the isotropic and anisotropic
magnetic behavior found in the different films. Amorphous and
hexagonal (non-cubic) substrate surfaces promote polycrystalline
magnetite films and isotropic magnetic behavior, whereas
(001)-oriented (cubic) substrate surfaces provide
epitaxial (001)-oriented magnetite films with an effective
fourfold (biaxial) magnetic anisotropy. In order to understand the rest of the
observed features, i.e. the particular orientation of the easy axes, the changes in the out-of-plane magnetization or the evolution of the Verwey temperature and the changes in remanence and coercivity with temperature  it is necessary to introduce the effects of the natural
growth defects,(i.e., APBs), as well as thickness effects on (pinned) domain wall dynamics.

Thus we suggest that the magnetic features of IR-PLD grown pure magnetite
films have three origins which are not independent of each other:

\begin{itemize}
    \item Substrate-control of magnetic symmetry, resulting in the isotropic magnetic behavior in the polycrystalline magnetite films and the fourfold magnetic anisotropy in the (001)-oriented epitaxial magnetite films.
    \item Antiphase boundaries (APBs)-control $T_{\rm V}$ and promote a magnetic symmetry along the $\langle100\rangle$ directions, i.e., rotated $45^{\circ}$ with respect to the bulk projected directions.
    \item Dimensionality-control of coercive field, following an inverse law dependence with the film thickness.
\end{itemize}

We start by discussing the role of antiphase domain boundaries.

\subsection{Antiphase Boundaries}
Observations of high coercivities, high saturation
fields, out-of-plane
magnetization~\cite{margulies1997origin}, superparamagnetism in
ultrathin films~\cite{voogt1998superparamagnetic}, biaxial
anisotropy induced by growth of magnetite films on MGO stepped
substrates~\cite{mcguigan2008plane}, as well as pinned magnetic
domain structures~\cite{wei2006influence} have all been attributed
to antiphase domain boundaries (APBs)~\cite{margulies1997origin}.

An antiphase domain boundary (APB) in a magnetite film appears
when two islands of magnetite, separated by a non-integer multiple
of the unit cell, coalesce. In such case, while the atomic lattice
is continuous across the boundary, the cations order is disrupted.
The presence of APBs in magnetite films has been extensively
investigated in films grown on
MgO(001)~\cite{celotto2003characterization,eerenstein2003diffusive},
which have an almost perfect lattice mismatch but with different
structures (spinel/cubic), and more recently on MAO
(111)~\cite{mckenna2014atomic} which share the same structure
(spinel/spinel) but have an small lattice mismatch (3$\%$). In both cases,
magnetite films show APBs. Moreover, it was found that the density
of APBs is related to the deposition temperature, and can be
decreased by post-deposition annealing. Under the same temperature
growth conditions, the average domain size within the APB
increases with the square root of the thickness, i.e., APBs
density decreases following an inverse law with the square root of
the thickness~\cite{celotto2003characterization}. The increase in
domain size and the decrease in the number of boundaries is a
consequence of the APBs annealing out of the films with time at
high temperature during growth and during
annealing~\cite{eerenstein2003diffusive}, as expected from
diffusive antiphase boundary coarsening
theory~\cite{allen1979microscopic}.

The effective magnetic easy axes found in the epitaxial films of
this study, i.e., along the in-plane $\langle100\rangle$
directions, can be correlated with the directionality and the
magnetic coupling of the APBs formed during
growth~\cite{celotto2003characterization}. APB shifts can be
formed based on the different translation and rotation symmetry
respectively for the case of the first Fe$_3$O$_4$ monolayer and
the (001)-oriented surface. The influence of APBs on magnetic
properties comes from the existence of specific geometries of the
Fe-O-Fe arrangements not present in perfect bulk magnetite. The
nature of magnetic coupling across APBs can be either
antiferromagnetic or ferromagnetic, as recently show by
atomic-resolution transmission electron microscopy and
differential phase contrast imaging measurements~\cite{chen2018}.
For example, a given fraction of APBs comprises aligned Fe$^{\rm
+3}$-O-Fe$^{\rm +3}$ bonds (see Fig.1 of
ref.~\cite{celotto2003characterization}) which are known to create
extremely strong antiferromagnetic superexchange interaction, resulting in a
strong antiferromagnetic coupling along the $\langle110\rangle$
magnetite surface directions. This might hinder the magnetic
orientation along this direction, even in the case that it would
correspond with the expected (bulk) anisotropy direction. In
contrast, $\langle100\rangle$ APBs are expected to couple
ferromagnetically. This might favor the magnetic orientation
along $\langle100\rangle$ directions.

Further effects of these structural defects on the magnetic and
transport properties of magnetite films are the high fields
required to saturate the magnetization~\cite{margulies1997origin}
and the large magnetoresistive effects~\cite{mcguigan2008plane}.
The presence of APBs is the origin of rotation of the spins in the
material and the formation of a complicated spin structure, due to
the strong antiferromagnetic coupling at APBs and with a partial
out-of-plane spin orientation around them, which depends on the
external field~\cite{niesenPRB2003}. Therefore, a smaller density
of APBs will result in a decrease of the effective out-of-plane
spin orientation.

The density and directionality of APBs have been analyzed
extensively by Celotto and coworkers for different magnetite thicknesses
and thermal treatments by using rose diagrams from image analysis
on dark field images~\cite{celotto2003characterization}. The
density of APBs (domain size) decreases (increases) with film
thickness and with increasing growth or annealing temperature. In
turn, the directionality of the APBs is predominantly oriented
close to $\pm\left[100 \right]$ and $\pm\left[010 \right]$
directions. This would make easier the magnetic
orientation along this direction. Remarkably, the APBs
directionality depicted in the rose diagrams of Fig.4 in
ref.~\cite{celotto2003characterization} closely resemble the
polar-plots of magnetic properties discussed and presented
previously (see Fig.~\ref{fig:PolarPlots} and
Fig.~\ref{fig:Hystpolar2}).

We thus suggest that the epitaxial films (i.e., those grown on
STO:Nb, STO, MAO, and MgO) present substantial densities of APBs
oriented as reported in ref.~\cite{celotto2003characterization}.
Within this scenario, APBs may act as pinning centers making
harder the magnetic orientation along specific directions, and
making easier the most distant to the latter. In this sense, the
effective hard axis directions lie along the $\langle110\rangle$
directions, from antiferromagnetic coupling between APBs, whereas
the effective easy axes are aligned along the $\langle100\rangle$
directions.

The presence and inferred evolution of APBs explains the
dependence of our ICEMS observations with thickness.
All films investigated by ICEMS
present an out-of-plane perpendicular component,
which has not been detected by MOKE. This out-of-plane
component is presumably originated
from APBs in the films wiht a local anti-ferromagnetic
ordering~\cite{niesenPRB2003,juan2019}, which have extremely
high coercive fields \cite{niesenPRB2003} and are thus not
detectable with our MOKE
setup. This component is more prominent (lower $x$) for the thinner films
(Fig.~\ref{fig:ICFMS_MgO}). So thinner films should have a somewhat higher APBs density.
This is in line with the known behavior of APBs, which is reduced for samples annealed at high temperature: as all the films were grown at 750~K and at the same rate, the thicker films are kept at high temperature for longer times.

\subsection{Reversal processes}
Magnetization reversal is determined by the
film microstructure (morphology, roughness,
defects density), and by the magnetic properties including the
anisotropy. The basic magnetization processes in ferromagnetic systems
are magnetization rotation, nucleation of inverted magnetic
domains and motion of magnetic domains walls.

In the case of magnetic film with well defined magnetic
anisotropies, coming from
a saturated state and decreasing the applied field strength, the
magnetization rotates in order to be aligned with the
magnetization easy axis direction. At zero field the magnetization
remanence depends on the angle between the direction of the last
external field used and the anisotropy axis. From the remanent
state, upon reversing the field orientation, the magnetization
rotates until at an specific field a reversed magnetic
domain nucleates, usually at a low coordination site. The reversal then
continues by a further propagation of the magnetic domain wall,
driven by the pressure exerted by the external field. Within this
scenario, coercivity is a property related to the rate at which
magnetic relaxation between the remanent and demagnetized states
takes place. The relaxation process involves displacements of
magnetic domain walls. For instance, APBs can act as pinning centers
for the domain wall
movement hardening the reversal.

Simple models based on the energy stored in a domain wall have
been proposed in order to analyze the dependence of $H_{\rm C}$
with the thickness of a FM film $t_{\rm FM}$. This energy arises,
among others, from exchange, anisotropy, and magnetostatic
contributions, and depends on the type of wall under motion (i.e.,
Bloch or Neel type). In particular, while in both cases the models
predict that the coercivity scales with the saturation
magnetization,  in the case of Neel wall motion the coercivity is
proportional to $t_{\rm FM}$ whereas for Bloch wall motion it is
inversely proportional to $t_{\rm FM}$. Neel walls are common in
(ultrathin) films where the exchange length, c.a. few nm in FM, is
very large compared to the thickness. For the case of our films,
with thicknesses between 30 to 450~nm, Bloch type (bulk-like)
domain walls must be considered. The inverse law with thickness
predicted by the bulk-like model has been already proved
experimentally in FM
films~\cite{min1998thickness,camarero2000thickness}.

In the case of the films studied, apart from the weak expected
evolution of the density of APB defects with the thickness
inferred from ICESM, the
rest of structural parameters are similar between each of the
epitaxial films. In fact, all have well-defined fourfold-symmetric magnetic
anisotropy (which is however rotated 45$^{\circ}$ from the projected bulk
directions) and similar values of crystal grain size (40 to
60~nm), strain ($<1\%$), and surface roughness ($<1$~nm) on flat
areas). Therefore the $1/t_{\rm Fe_3O_4}$ law found in these
films, independently of the substrate used, can be explained by the
general behaviour of thin layers with Bloch type domain walls, i.e.,
bulk-like layers, mediated by the APBs. This is also supported by
the thickness dependence of the remanent magnetization. The
thinnest samples show an \textit{M$_R$/M$_S$} ratio close to 1,
ratio which decreases as the thickness increases.
The APB-induced unexpected magnetic anisotropy also diminishes,
as the density of APBs is somewhat reduced with thickness.

\begin{figure}[h!]
    \begin{center}
        \includegraphics[width=1.0\textwidth]{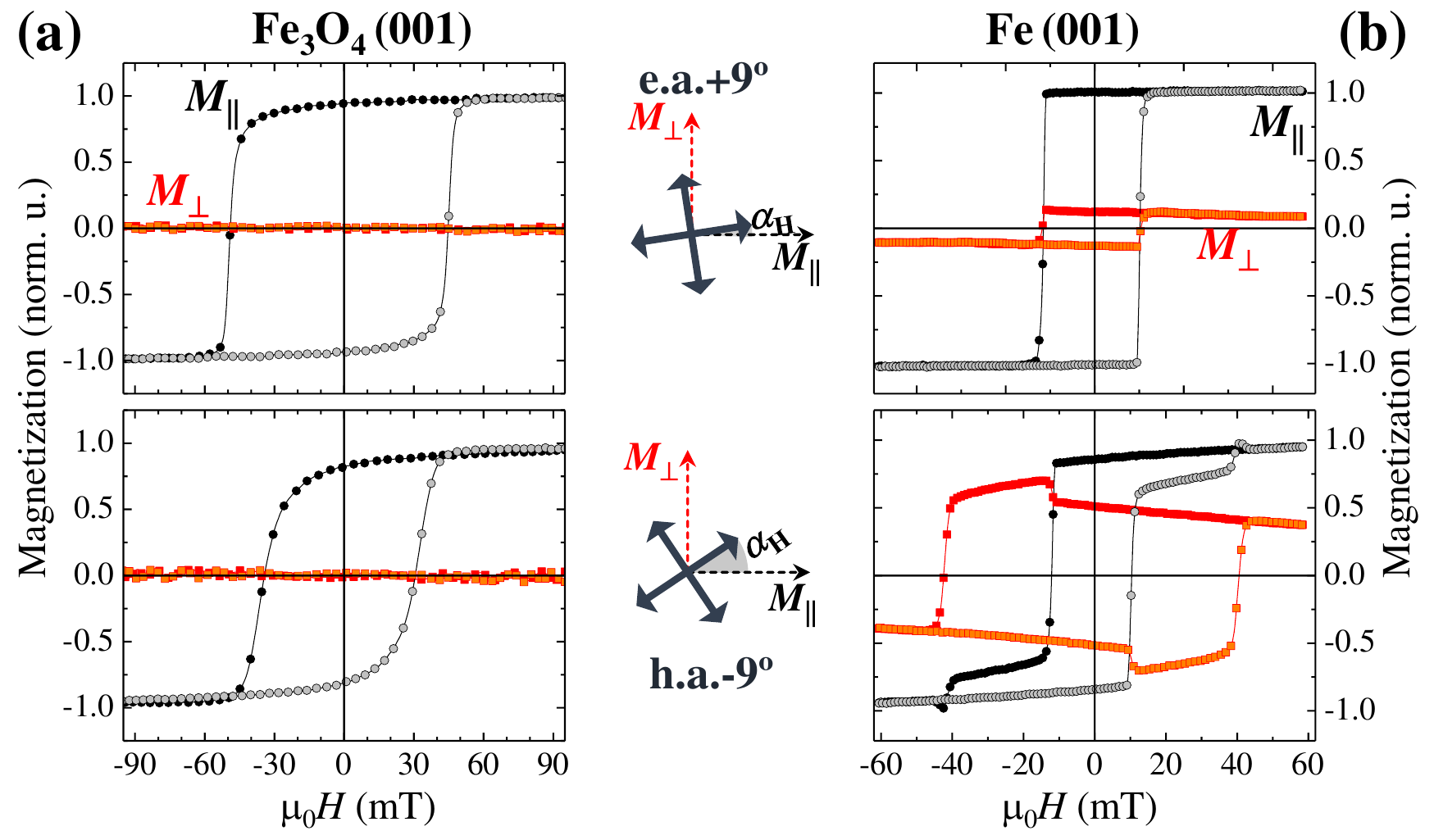}
        \caption{Selected vectorial-resolved hysteresis loops of
        films with fourfold magnetic symmetry acquired at
        indicated angles $\alpha_{\rm H}$: Fe$_3$O$_4$(001) (a);
        Fe(001) (b). The corresponding measure geometries, where
        the e.a. directions (magnetization components) are
        indicated by thick continuous (thin dashed) arrows are
        shown in the middle. The $M_{\rm \parallel}(H)$ and
        $M_{\rm \perp}(H)$ loops are represented by circles and
        squares, respectively. The two branches have been depicted
        with different filled symbols to clarify the evolution of
        the magnetization.} \label{fig:4foldCompare}
    \end{center}
\end{figure}

Further proof that the origin of observed magnetic anisotropy in
the (001)-oriented films is related to the
morphological distribution of the APB defects is the fact that the
associated phenomena is different from that usually found in
epitaxial (001)-oriented cubic films. Fig.~\ref{fig:4foldCompare}
compares representative vectorial-resolved measurements of the
epitaxial films (left graphs) with the ones of a cubic Fe(001)
film (right) acquired with the field oriented nearby the
characteristic easy (top) and hard (bottom) directions. In
general, both systems present fourfold symmetry, i.e., the
magnetic properties are repeated every 90$^{\rm \circ}$, but the
reversal pathways are quite different. For example, for the
Fe(100) film, the reversal is characterized by a strong angular
dependence in the number of irreversible
transitions~\cite{jimenez2014vectorial}. In particular, when the
field orientation is closed to the easy directions, an
irreversible transition (corresponding to nucleation and
propagation of 180-oriented domain walls) is observed, whereas two
irreversible transitions (related to nucleation and propagation of
90-oriented domain walls) are found close to the hard directions.
Note that this is also easier to identify in the $M_{\rm
\perp}$($H$) loop. In clear contrast, the vectorial-resolved loops
of any of the epitaxial magnetite films investigated show just one
irreversible transition and only in the $M_{\rm \parallel}$($H$)
loop, being almost negligible the $M_{\rm \perp}$($H$) loop. The
latter could be explained by opposite magnetization rotation
pathways, i.e., above and below the field direction, taking place
before and after the sharp transition. This would cancel out the
$M_{\rm \perp}$ signal at any field, i.e., $M_{\rm
\perp}(H)\approx 0$, as observed experimentally.

\section{Conclusions}
\label{sec:Conclussions} High quality stoichiometric magnetite
(Fe$_3$O$_4$) films grown by infrared pulsed laser deposition
(IR-PLD) on different substrates (i.e., SrTiO$_3$(001),
MgAl$_2$O$_4$(001), MgO(001), Al$_2$O$_3$(0001) and amorphous
Si/SiO$_2$) have been investigated in order to study the influence
of the substrate, orientation, and thickness on their magnetic
behavior.

All films consist of nanocrystalline stoichiometric magnetite
with very small strain ($< 1\%$) and a Verwey transition
($T_{\rm V}$) between 110-120~K, i.e., close to the transition temperature
of bulk magnetite
(125~K). $T_{\rm V}$ depends on microstructure
and thickness, increasing as the thickness increases. Room
temperature angular-dependent measurements reveal isotropic
behavior for magnetite films grown on Al$_2$O$_3$(0001) and
Si/SiO$_2$, whereas an in-plane fourfold symmetry magnetic
behavior for all films grown on (001)-oriented surfaces, and with
the easy axes lying along the Fe$_3$O$_4$ [010] and [100]
directions, i.e., rotated with respect to the bulk projected directions.
Remarkably, the fourfold magnetic symmetry is shown even in
400~nm thick films. In turns, the coercive field ($H_{\rm C}$)
depends on microstructure and film thickness. The largest
(lowest) $H_{\rm C}$ value has been found for the thinner film
grown on a single crystal SrTiO$_3$(001) (amorphous Si/SiO$_2$)
surface. Moreover, the coercivity follows an inverse law with film
thickness, as predicted with a simple bulk-like model.

These results demonstrate that it is possible to artificially
control the magnetic behavior of stoichiometric IR-PLD grown
Fe$_3$O$_4$ films by exploiting substrate-induced anisotropy and
thickness-controlled coercivity, helping paving the way to
incorporate magnetite in future magnetic-based applications.

\section{Acknowledgements}
This research was supported by the Regional Government of Madrid
through Projects S2013/MIT-2850 (NANOFRONTMAG) and P2018/NMT-4321
(NANOMAGCOST) and by the Spanish Ministry of Economy and
Competitiveness (MINECO) through Projects FIS2016-78591-C3-1-R,
CTQ2016-75880-P, and MAT2017-89960-R. M. O. thanks CSIC for
contract and E.R. thanks MINECO for the tenure of a Ram\'{o}n y
Cajal contract (No. RYC-2011-08069). IMDEA Nanoscience is
supported by the 'Severo Ochoa' Programme for Centres of
Excellence in R$\&$D, MINECO [grant number SEV-2016-0686].

\end{document}